\documentclass[aps,pra,twocolumn,showpacs,superscriptaddress,groupedaddress]{revtex4}

\usepackage{graphicx}
\usepackage{bm}
\usepackage{amssymb}
\usepackage{amsmath}
\usepackage{braket}

\usepackage{hhtensor}
\usepackage{algpseudocode}
\usepackage{algorithm}
\usepackage{mathtools}
\usepackage{blkarray}
\usepackage{multirow}
\usepackage{hhline}

\usepackage[colorlinks]{hyperref}

\newcommand{\Utarget}{U_{\text{target}}}
\newcommand{\I}{\mathbb{I}}

\newcommand{\real}{\mathbb{R}}
\newcommand{\Lone}[1]{\mathbf{L}_1\left(\real,\real^{#1}\right)}
\newcommand{\trace}{\operatorname{Tr}}

\newcommand{\conv}{\star}

\newcommand{\re}{\operatorname{Re}}
\newcommand{\im}{\operatorname{Im}}
\newcommand\ii{\mathrm{i}}
\newcommand{\expect}{\mathbb{E}}
\newcommand{\defeq}{\mathrel{\mathop:}=}

\newcommand{\eq}[1]{\hyperref[eq:#1]{(\ref*{eq:#1})}}
\renewcommand{\sec}[1]{\hyperref[sec:#1]{Section~\ref*{sec:#1}}}
\newcommand{\app}[1]{\hyperref[app:#1]{Appendix~\ref*{app:#1}}}
\newcommand{\fig}[1]{\hyperref[fig:#1]{Figure~\ref*{fig:#1}}}
\newcommand{\thm}[1]{\hyperref[thm:#1]{Theorem~\ref*{thm:#1}}}
\newcommand{\lem}[1]{\hyperref[lem:#1]{Lemma~\ref*{lem:#1}}}
\newcommand{\cor}[1]{\hyperref[cor:#1]{Corollary~\ref*{cor:#1}}}
\newcommand{\defn}[1]{\hyperref[def:#1]{Definition~\ref*{def:#1}}}
\newcommand{\alg}[1]{\hyperref[alg:#1]{Algorithm~\ref*{alg:#1}}}

\begin{document}


\title{Accounting for Classical Hardware in the Control of Quantum Devices}

\author{I.N. Hincks}
\affiliation{Department of Applied Math, University of Waterloo, Waterloo, ON, Canada}
\affiliation{Institute for Quantum Computing, Waterloo, ON, Canada}

\author{C.E. Granade}
\affiliation{Department of Physics, University of Waterloo, Waterloo, ON, Canada}
\affiliation{Institute for Quantum Computing, Waterloo, ON, Canada}

\author{T.W. Borneman}
\affiliation{Department of Physics, University of Waterloo, Waterloo, ON, Canada}
\affiliation{Institute for Quantum Computing, Waterloo, ON, Canada}

\author{D.G. Cory}
\affiliation{Department of Chemistry, University of Waterloo, Waterloo, ON, Canada}
\affiliation{Institute for Quantum Computing, Waterloo, ON, Canada}
\affiliation{Perimeter Institute for Theoretical Physics, Waterloo, ON, Canada}
\affiliation{Quantum Information Science Program, Canadian Institute for Advanced Research, Toronto, ON, Canada}

\date{\today}
\pacs{}
\begin{abstract}
High fidelity coherent control of quantum systems is critical to building quantum devices and quantum computers.
We provide a general optimal control framework for designing control sequences that account for hardware control distortions while maintaining robustness to environmental noise.
We demonstrate the utility of our algorithm by presenting examples of robust quantum gates optimized in the presence of nonlinear distortions.
We show that nonlinear classical controllers do not necessarily incur additional computational cost to pulse optimization, enabling more powerful quantum devices.
\end{abstract}
\maketitle


The ability to coherently control the dynamics of quantum systems 
with high fidelity is a critical component of the development of 
modern quantum devices, including quantum computers \cite{ladd_quantum_2010}, 
actuators \cite{hodges_universal_2008,borneman_parallel_2012}, 
and sensors \cite{cappellaro_entanglement_2005,mamin_nanoscale_2013,taylor_high-sensitivity_2008} 
that push beyond the capabilities of classical computation and metrology.
In recent years, quantum computation has
presented a compelling application for quantum control, as high-fidelity
control is essential to implement quantum information processors that
achieve fault-tolerance  
\cite{gottesman_introduction_2009,fowler_high-threshold_2009,gutierrez_approximation_2013}. 

The performance of numerically optimized quantum gates in 
laboratory applications strongly depends on the accuracy of 
the system model used to approximate the response of the 
experimental system to the applied control sequence.
Here we develop a general framework whereby classical control
hardware components are modelled explicitly, such that their effect on a
quantum system can be computed and compensated for using numerical optimal 
control theory (OCT) \cite{pontryagin_mathematical_1987} algorithms to optimize control
sequences. Control sequences designed using OCT algorithms, such as the
GRadient Ascent Pulse Engineering (GRAPE)~\cite{khaneja_optimal_2005} algorithm,
can be made robust to a wide variety of inhomogenities, pulse errors and noise processes \cite{kobzar_exploring_2012,koroleva_broadband_2013,borneman_application_2010}.
These methods are also easily extended 
~\cite{mandal_axis-matching_2013,mandal_direct_2014,schulte-herbruggen_optimal_2011,goerz_optimal_2014}
to other applications and may be integrated into 
other protocols~\cite{egger_adaptive_2014}. 
Recently, it was demonstrated how a model of linear distortions of the 
control sequence, such as those arising from finite bandwidth of the classical 
control hardware, may also be integrated into OCT algorithms
~\cite{spindler_shaped_2012,borneman_bandwidth-limited_2012,motzoi_optimal_2011}.

Here, we improve and generalize those results beyond linear kernels to 
any operation that smoothly maps a list of control steps to a classical
field seen by the quantum system.
Importantly, our framework naturally allows for robustness against 
uncertainties and errors due to classical control hardware.
We begin developing our method generally, without making 
assumptions about the device of interest, so that our results may be broadly
applicable to a wide range of quantum devices.
We briefly discuss how our theory is easily applied to any linear distortion,
and then in more detail, demonstrate  with numerics how nonlinearities 
in control hardware, such as those found in strongly-driven 
superconducting resonators used for pulsed electron spin resonance (PESR) \cite{benningshof_superconducting_2013,malissa_superconducting_2013,sigillito_fast_2014}, 
may be included in OCT algorithms. 


With this goal in mind, we briefly review the problem of controlling a
quantum system~\cite{dalessandro_introduction_2007}.
Given a system Hamiltonian
\begin{equation}
	H(t)=H_0+\sum_{l=1}^L q_l(t)H_l
	\label{eqn:hamiltonian}
\end{equation}
where $H_0$ is the internal Hamiltonian and $\{H_l\}_{l=1}^L$ are the control
Hamiltonians, how do we choose the envelopes
$\{q_l(t)\}_{l=1}^L$ such that at time $T$ we effect the total unitary $\Utarget$?
It will be clear that the framework we construct will be compatible with not only
this specific unitary control problem, but all similar problems such as 
state to state transfers, expectation values over static distributions,
open system maps, etc.

\begin{center}
\begin{figure}
	\includegraphics[width=0.4\textwidth]{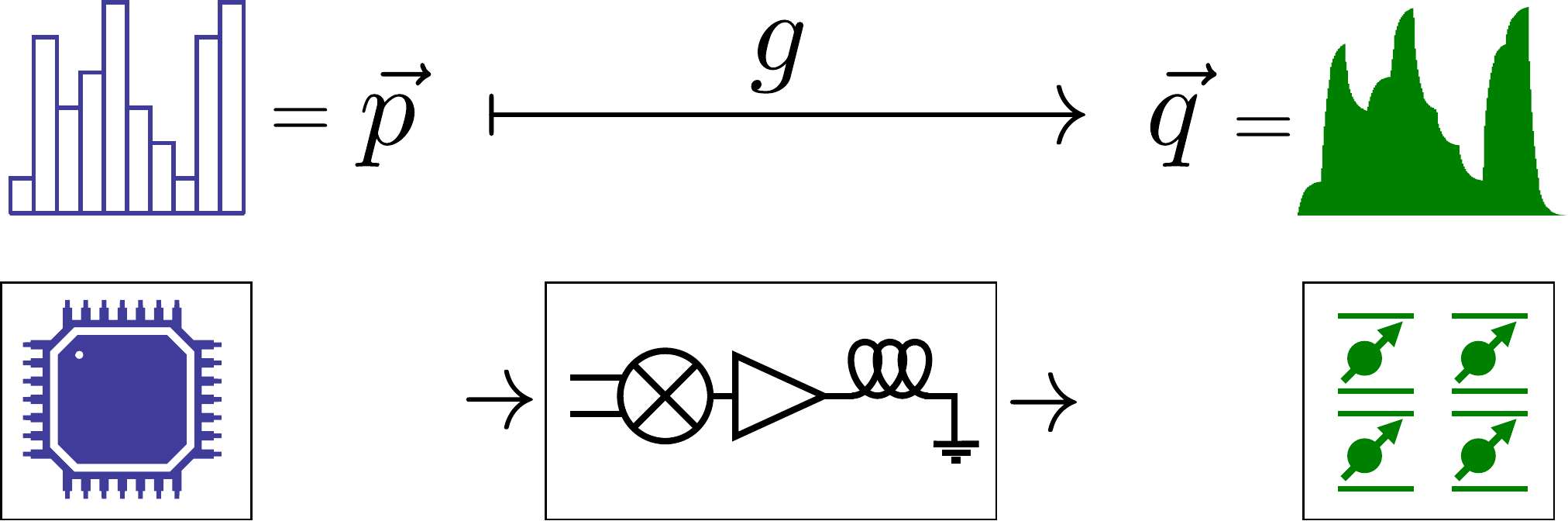}
	\caption{
	A cartoon depicting the action of the distortion operator $g$ on the input pulse $\vec{p}$.
	}
	\label{fig:framework}
\end{figure}
\end{center}

The functions $\{q_l(t)\}_{l=1}^L$ seen by the quantum system represent a distorted
version of what was input to the classical hardware. 
Since we are ultimately interested in doing numerics, we begin by 
discretizing the time domain and therefore model all relevant hardware 
by what we will call a \textit{discretized distortion operator}.
This is a function $g:\real^N\otimes\real^K\rightarrow\real^M\otimes\real^L$
which takes an input pulse sequence, $\vec{p}$, with some associated time step
$dt$, and outputs a distorted version of the pulse, $\vec{q}=g(\vec{p})$, with 
an associated time step $\delta t$.
$\vec{p}$ is the pulse as generated by the experimenter's computer, and $\vec{q}$ is
the pulse generating the Hamiltonian seen by the quantum system, as illustrated 
in Figure~\ref{fig:framework}.

The integers $N$ and $M$ are the number of input and output time steps respectively,
and $K$ and $L$ are the number of input and output control fields respectively.
In the case of quadrature control of a qubit, $K=L=2$.
We omit subscripts on the time steps $dt$ and $\delta t$ for
notational simplicity; uniform time discretization is not required.
Typically, we will have $\delta t < dt$ to allow for an accurate simulation 
of the quantum system.
The condition $M\cdot\delta t = N\cdot dt$ need not hold,
for example, $M\cdot\delta t > N\cdot dt$ will be useful
when the distortion has a finite ringdown time.

The discretized distortion operator $g$ will often derive from a 
\textit{continuous distortion operator} 
$f:\Lone{K}\rightarrow\Lone{L}$ which takes a continuous input pulse $\alpha(t)$ 
and outputs a distorted pulse $\beta(t)=f[\alpha](t)$.
The discretized version is obtained by composing $f$ on either side
by a discretization and dediscretization operator, $g= f_1\circ f \circ f_2$.

We can incorporate the distortion operator $g$ into 
standard techniques from optimal control theory.
In particular, consider the unitary objective function,
\begin{equation}
	\Phi[\vec{q}]=
		\left| \trace\left( 
			\Utarget^\dagger \prod_{m=1}^M e^{-i\delta t(H_0+\sum_{l=1}^Lq_{m,l}H_l)} 
		\right) \right|^2/d^2,
\end{equation}
used in the GRAPE algorithm~\cite{khaneja_optimal_2005}.
Penalties can be added to this basic objective function in order to demand
that the solution admit certain properties. For instance, penalty
functions have been used to ensure robustness to control noise
and limited pulse fluence
~\cite{hocker_characterization_2014,skinner_reducing_2004,koroleva_broadband_2013,kobzar_exploring_2008} or
to ensure that undesired subspaces are avoided \cite{palao_protecting_2008,muller_optimizing_2011}.

Now we include the effect of our hardware by modifying the objective function
to compose with the distortion operator,
\begin{equation}
	\Phi_g[\vec{p}]=\Phi\circ g(\vec{p}).
\end{equation}
Using the multivariable chain rule, we compute the gradient of $\Phi_g$ to be
\begin{align}
	& \nabla_{\vec{p}}(\Phi_g)=\nabla_{g(\vec{p})}(\Phi) \cdot J_{\vec{p}}(g) \\
	& \left[ J_{\vec{p}}(g) \right]_{m,l,n,k}
		= \frac{\partial g_{m,l}}{\partial p_{n,k}}
\end{align}
where the dot represents a contraction over the indices $m$ and $l$, 
and where $J_{\vec{p}}(g)$ is the Jacobian of $g$ at $\vec{p}$.
Though evaluating $\nabla_{g(\vec{p})}(\Phi)$ naively would require
simulating the action of $M\times L$ pulses, the
GRAPE algorithm \cite{khaneja_optimal_2005} provides an expression for this
gradient in terms of the timestep unitaries that are already computed,
\begin{equation}
    \frac{\partial \Phi}{\partial q_{m,l}} = -2\re\left[
        \braket{P_m | \ii \delta t H_l X_m}
        \braket{X_m | P_m}
    \right],
\end{equation}
where $P_m := \left(\prod_{i=m+1}^{M} U_i^\dagger\right)U_{\text{target}}$,
$X_m := \prod_{i=m}^{1} U_i$ and where $U_i(\vec{q}) =
\exp(-\ii\delta t[ H_0 + \sum_{l=1}^L q_{m,l} H_l ])$.
Therefore if we can compute the Jacobian $J_{\vec{p}}(g)$, we can then 
compute the total gradient of $\Phi$.
The rest of the algorithm follows as described in the original GRAPE~\cite{khaneja_optimal_2005}.

Since the cost of evaluating $g$ will typically not grow more than
polynomially with the number of qubits,
the computational cost of the optimization effectively remains unchanged 
from standard GRAPE, as it is still dominated by the cost of computing 
the $M$ matrix exponentials.


Our first example is the continuous distortion operator given by the convolution 
with an $L\times K$ kernel $\phi(t)$,
\begin{equation}
	\beta(t)=f(\alpha)(t)=(\phi\conv\alpha)(t)=\int_{-\infty}^\infty \phi(t-\tau)\cdot\alpha(t) d\tau.
\end{equation}
The convolution kernel $\phi$ models any distortion that can be described by a
linear differential equation, such as a simple exponential rise time, control line crosstalk, 
or the transfer function of the control hardware 
\cite{spindler_shaped_2012,rife_transfer-function_1989,borneman_bandwidth-limited_2012,barends_superconducting_2014}.
We compute the discretized distortion operator to be
\begin{equation}
	q_{m,l}=\sum_{n=1,k=1}^{N,K} \left(
		\int_{(n-1)dt}^{n dt} \phi_{l,k}((m-1/2)\delta t-\tau) d\tau
	\right) p_{n,k}.
\end{equation}
where we see that it acts as a linear map,
\begin{equation}
	\vec{q}=g(\vec{p})=\tilde{\phi}\cdot\vec{p},
\end{equation}
where we are contracting over the $n$ and $k$ indices with the components of
the tensor $\tilde{\phi}$ given by the integrals
\begin{equation}
	[\tilde{\phi}]_{m,l,n,k}=\int_{(n-1)dt}^{n dt} \phi_{l,k}((m-1/2)\delta t-\tau) d\tau.
	\label{eqn:convolution_component}
\end{equation}
The Jacobian matrix is simply given by $J_{\vec{p}}(g)=\tilde{\phi}$ which is 
independent of the pulse $\vec{p}$.

\begin{center}
\begin{figure}
	\includegraphics[width=0.45\textwidth]{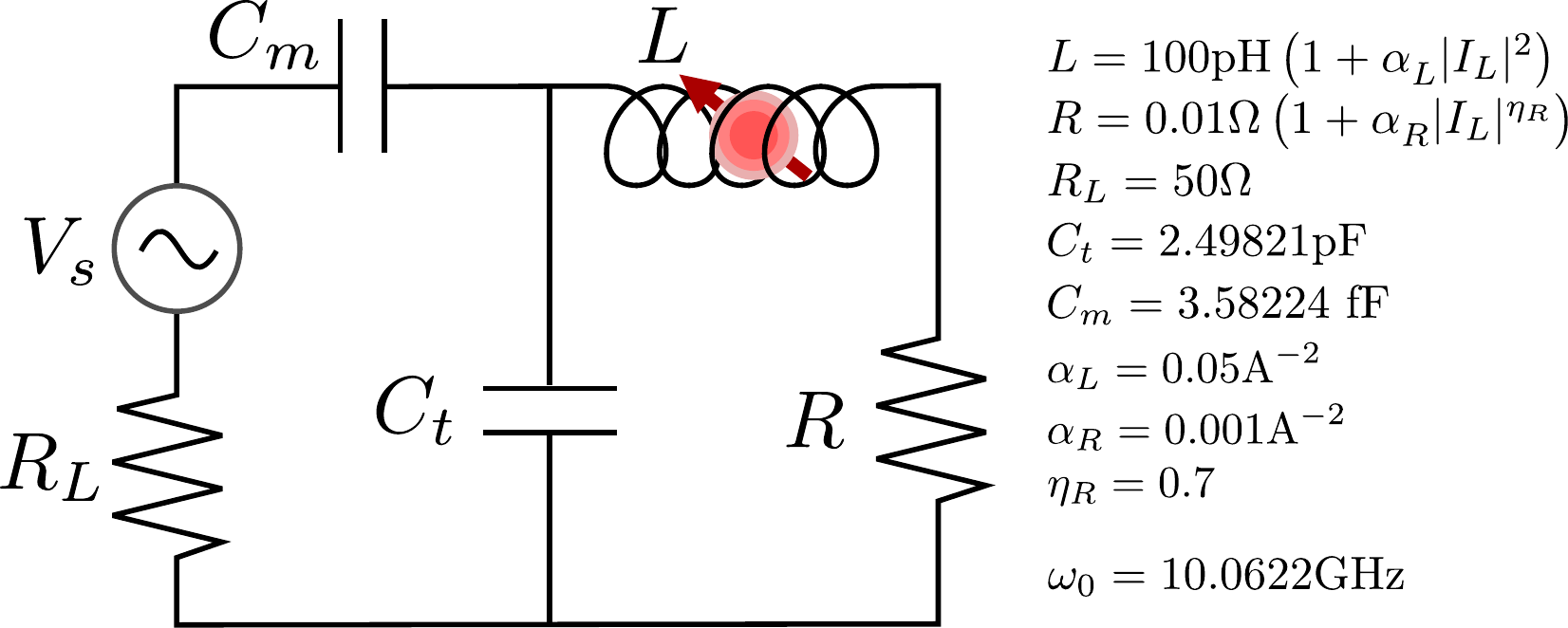}
	\caption{
	A quantum system being controlled by the magnetic field produced by the
	inductor of a nonlinear resonator circuit.
	The ideal voltage source $V_s(t)$ is specified by the input undistorted 
	pulse $\vec{p}$, and the resulting current through the inductor, $I_L(t)$,
	is computed.
	The inductance and the resistance are both functions of the current 
	passing through them.
	The form of the nonlinearity is chosen to be consistent with kinetic 
	inductance.
	}
	\label{fig:rlc}
\end{figure}
\end{center}

As a more involved example we consider a quantum system being controlled 
by a tuned and matched resonator circuit~\cite{barbara_phase_1991} with 
nonlinear circuit elements (Figure~\ref{fig:rlc}).
Nonlinear resonators are used in a variety of applications, including 
superconducting qubits for quantum information processing 
\cite{schoelkopf_wiring_2008}, microwave kinetic inductance detectors 
for astronomy \cite{day_broadband_2003}, and increasing inductive
detection sensitivity in magnetic resonance~\cite{bachar_nonlinear_2012}.
Often, however, electronics controlling quantum systems are operated in
their linear regime to avoid complications resulting from nonlinearity 
~\cite{sigillito_fast_2014}. 
Avoiding nonlinearities requires reducing input power, leading to 
longer control sequences that reduce the number of quantum operations 
that can be performed before the system decoheres. 
Additionally, limiting input power removes the natural robustness of 
high-power sequences to uncertainties in the environment achieved by 
strongly modulating the quantum system 
\cite{fortunato_design_2002,pravia_robust_2003}. 

If the circuit were linear, the distortion could be modelled as a 
convolution $\phi\conv{}\!$ as discussed above.
However, with nonlinear circuit elements present we must numerically solve the 
circuit's differential equation every time we wish to compute the distorted pulse 
\cite{maas_nonlinear_2003}.

For concreteness, but with no loss of generality, we work with an 
on-resonance qubit system whose Hamiltonian in the rotating frame, after 
invoking the rotating wave approximation, is
\begin{equation}
	H=
	\frac{\delta\omega}{2}\sigma_z+
	(1+\kappa)\left(
		\frac{\omega_x(t)}{2}\sigma_x+\frac{\omega_y(t)}{2}\sigma_y
	\right)
	\label{eqn:qubit_hamiltonian}
\end{equation}
where $\delta\omega$ and $\kappa$ represent off-resonance and control
power errors, respectively.

\begin{center}
\begin{figure}
	\includegraphics[width=0.45\textwidth]{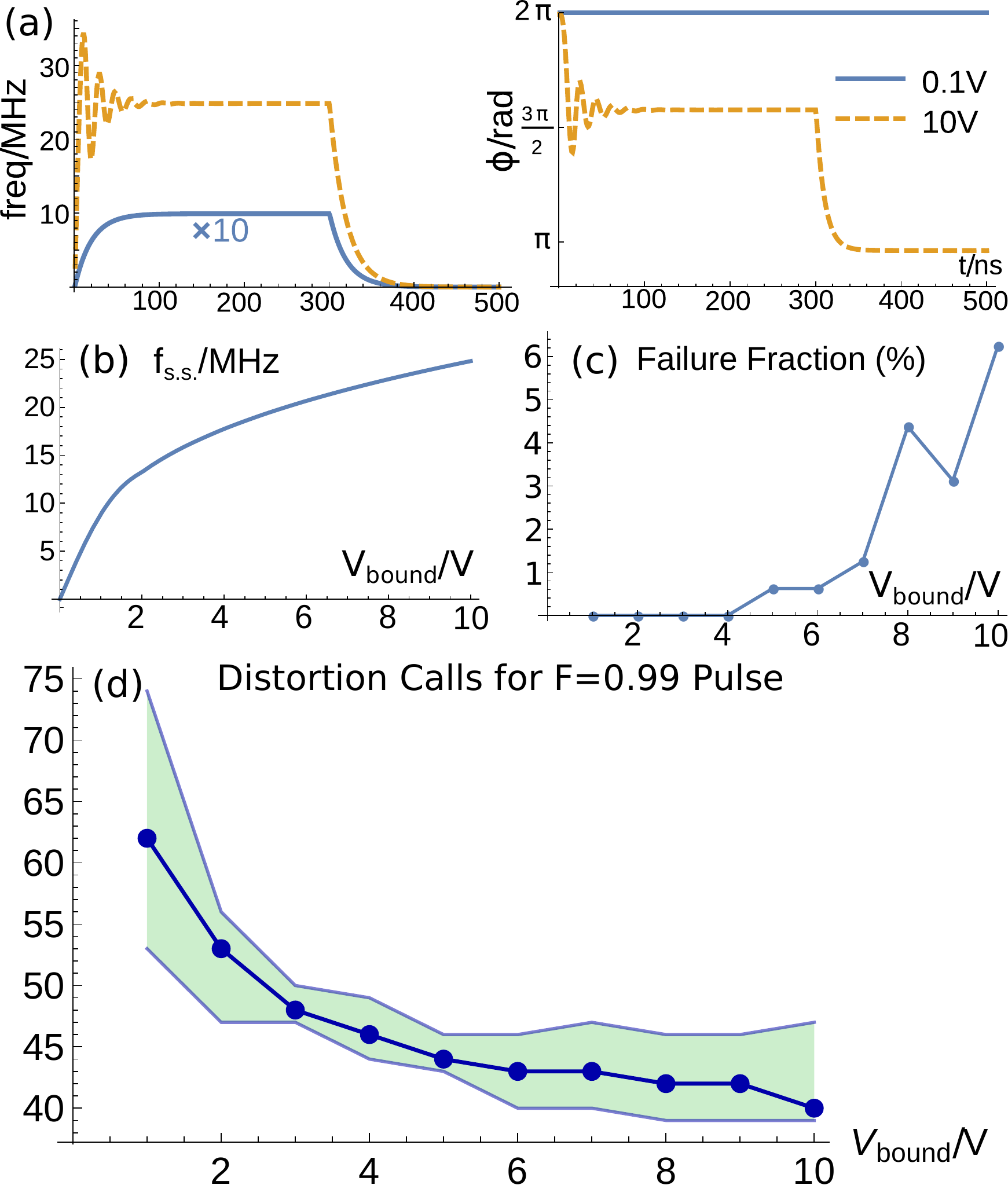}
	\caption{
	(a) Response from the same resonator to a square forcing term with 
	length $300$ns in both a linear (0.1V) and nonlinear (10V) regime.
	The amplitude of the 0.1V pulse is multiplied by $10$ to make it visible.
	(b) The steady state driving frequency as seen by the spins as a 
	function of the voltage input to the resonator.
	(c) Out of $160$ pulses searched for at each of $10$ voltage bounds, the fraction
	that failed to reach $F=0.99$ before the step size was effectively zero, and
	(d) the median number of calls made to the distortion function $g$ along with 
	the $16\%$ and $84\%$ quantiles during the gradient ascent for those pulses which 
	did reach $F=0.99$.
	}
	\label{fig:cost}
\end{figure}
\end{center}

The time evolution of the circuit shown in Figure~\ref{fig:rlc} is governed by the third 
order differential equation
\begin{equation}
	\frac{d}{dt}
	\begin{bmatrix}
		I_L \\ V_{C_m} \\ V_{C_t}
	\end{bmatrix}
	=
	\begin{bmatrix}
		-\frac{R}{L} & 0 & \frac{1}{L} \\
		0 & \frac{-1}{R_L C_m} & \frac{1}{R_L C_m} \\
		\frac{-1}{C_t} & \frac{-1}{R_L C_t} & \frac{1}{R_L C_t}
	\end{bmatrix}
	\begin{bmatrix}
		I_L \\ V_{C_m} \\ V_{C_t}
	\end{bmatrix}
	+
	\begin{bmatrix}
		0 \\ \frac{V_s(t)}{R_L C_m} \\ \frac{V_s(t)}{R_L C_t}
	\end{bmatrix}
	\label{eqn:resonator_de_lab}
\end{equation}
where the nonlinearities arise when the inductance, $L$, and resistance,
$R$, are functions of the current passing through them
\cite{maas_nonlinear_2003,tinkham_introduction_2004}.
In the case of kinetic inductance, these nonlinearities take on the form
\begin{align}
	L=L(I_L)=L_0(1+\alpha_L |I_L|^2) \nonumber \\
	R=R(I_R)=R_0(1+\alpha_R |I_R|^\eta) 
\end{align}
where $\alpha_L$, $\alpha_R$ and $\eta$ are constants~\cite{dahm_theory_1996,mohebbi_composite_2014}.
Kinetic inductance leads to a reduction in the circuit resonance frequency,
coupling, and quality factor with increasing power, as shown in Figure~\ref{fig:cost}(a-b).

Since our Hamiltonian in Equation~\ref{eqn:qubit_hamiltonian} is written in 
a frame rotating at the circuit resonance frequency in the linear-regime, 
it is convenient to write our differential equation in this frame.
To this end, with the differential equation~\ref{eqn:resonator_de_lab} shorthanded 
as $\dot{\vec{y}}(t)=B(\vec{y}(t))\vec{y}(t)+V_s(t)\vec{b}$, introduce the complex
change of variables $\vec{x}(t)=e^{-i\omega_0 t}\vec{y}(t)$.
In this new frame, since $B(\vec{y}(t))=B(\vec{x}(t))$, our dynamics become
\begin{align}
	\dot{\vec{x}}(t)&=\left(
		B(\vec{x}(t))-i\omega_0 \I
	\right)\vec{x}(t)
	+\tilde{V}_s(t)\vec{b} \nonumber \\
	&\equiv
	A(\vec{x}(t))\vec{x}(t)+\tilde{V}_s(t)\vec{b}
	\label{eqn:resonator_de}
\end{align}
where we have invoked the rotating wave approximation, and $\tilde{V}_s(t)$ is
the rotating version of $V_s(t)$.
Now the real and imaginary parts of the complex current in the rotating 
frame, $\tilde{I}_L(t)=e^{-i\omega_0t}I_L(t)$, are proportional via a geometric
factor to the control amplitudes appearing in the Hamiltonian,
\begin{equation}
	\omega_x(t)\propto\re[\tilde{I}_L(t)] ~\text{ and }~
	\omega_y(t)\propto\im[\tilde{I}_L(t)].
\end{equation}

To compute the distortion $\vec{q}=g(\vec{p})$ caused by the resonator, we set 
the circuit's input voltage $\tilde{V}_s(t)$ to be the piecewise constant function
with amplitudes coming from $\vec{p}$.
To improve stiffness conditions, a small finite risetime may be added to the forcing term 
$\tilde{V}_s(t)$, which is equivalent to adding a low-pass filter to the ideal voltage 
source in the circuit.
We can now solve the equations~\ref{eqn:resonator_de} for $\tilde{I}_L(t)$ using the
NDSolve function in Mathematica 10, interpolate the results, and resample at a rate $\delta t$ to 
determine the distorted pulse $\vec{q}$.

\begin{figure*}
    \centering
    \includegraphics[width=0.81\textwidth]{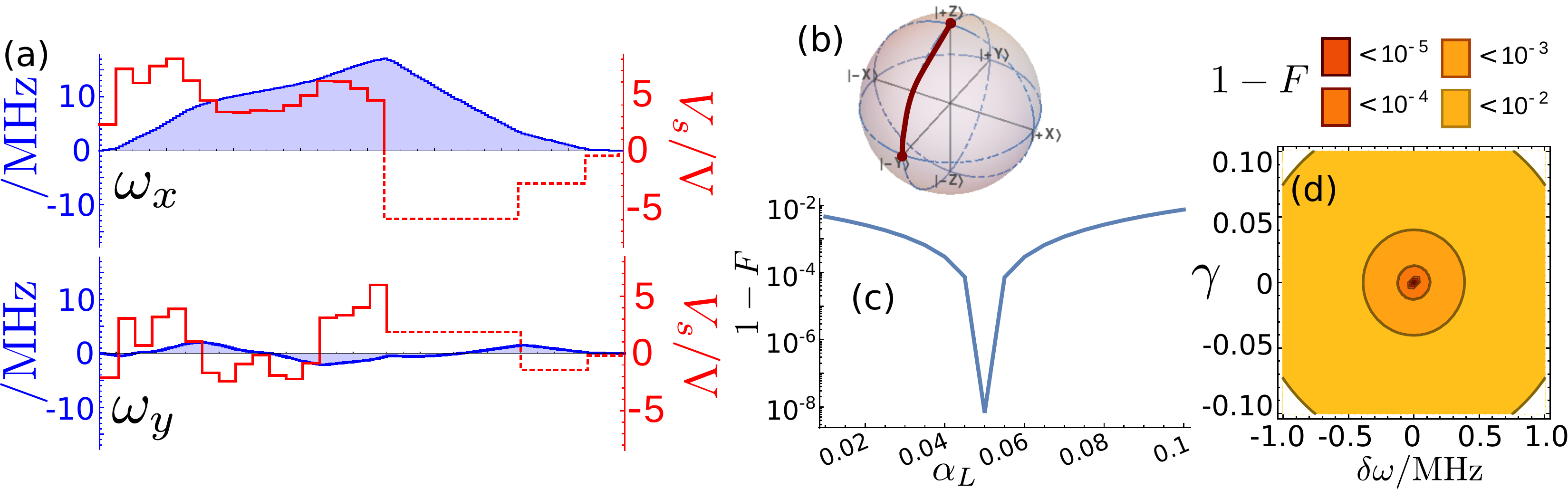}
    \caption{\label{fig:example-nonlin-pulse}
        (color online)
        (a) Example of a $\pi/2)_x$ pulse generated for the matched nonlinear resonator
        circuit.
        The driving term ($\vec{p}$) is shown in red, while the distorted pulse ($\vec{q}$) is shown in
        blue.
        The dashed segments are the ringdown compensation steps.
        (a) The trajectory of the state $\ket{0}$ under this pulse is shown on the Bloch sphere, and
        (c-d) the average fidelity is plotted for different values of $\alpha_L$, $\delta\omega$, and $\gamma$.
    }
\end{figure*}

Since our distortion is nonlinear, the Jacobian of $g$ will not be 
constant with respect to the input pulse $\vec{p}$.
However, we may compromise the accuracy of the Jacobian in favour of 
taking a larger number of ascent steps that are still generally uphill 
by considering only the Jacobian at the zero pulse,
\begin{equation}
	\frac{\partial g_{m,l}}{\partial p_{n,k}}\bigg|_{\vec{p}}
	\approx
	\left[
		g(\epsilon \vec{e}_{n,k})/\epsilon
	\right]_{m,l}.
	\label{eq:linear-approx}
\end{equation}
These quantities may be precomputed prior to gradient ascent and 
therefore only add a constant to the computation time.
Exact partial derivatives may be computed for a cost that scales as
$K\cdot N$ and whose implementation can be highly parallelized; see the 
Appendix for details.

In \fig{example-nonlin-pulse}, we show an example of a GRAPE-optimized
pulse for $U=\frac{\pi}{2})_x$, with the circuit of \fig{rlc} used as a distortion
operator. 
There are $16$ times steps of length $0.5$ns shown as a solid red step function.
The pulse has been made to be robust to static uncertainty in the Hamiltonian 
parameters $\delta\omega$ and $\gamma$ and the nonlinearity parameter $\alpha_L$.
Since the circuit has a high quality factor, it would take many times the length
of the pulse for the ringdown tail to decay to zero.
We therefore utilize an active ringdown suppression scheme with three compensation 
steps of lengths $4$ns, $2$ns, and $1$ns.
This is a generalization of ringdown suppression in linear 
circuits~\cite{borneman_bandwidth-limited_2012,mehring_high_1976,hoult_fast_1979}
and is discussed in detail in the Appendix.

The inclusion of a distortion operator causes an alteration to the control landscape 
which might be expected to make finding optimal solutions more expensive.
Therefore, a tradeoff between computational cost and gate time length might be anticipated.
We perform a numerical study to examine this relationship.

We bound the allowed input power to the resonator used by the GRAPE algorithm 
by $10$ different voltages, $1~\mathrm{V}$ to $10~\mathrm{V}$, where $1~\mathrm{V}$ is on the edge of the linear 
regime, and $10~\mathrm{V}$ is highly nonlinear.
In analogy to the numerical control landscape experiments performed in Reference~\cite{moore_relationship_2008},
for each of these bounds, we attempt to compute a fidelity $F=0.99$ $\frac{\pi}{2})_x$ 
pulse $160$ times, with a different random initial guess each time.
The total length of the pulse is set to $T_\text{pulse}=\frac{0.25}{f_\text{s.s.}}$
where $f_\text{s.s.}$ is the steady state driving frequency of the resonator at 
the corresponding voltage bound.
The number of time steps is held constant at $N=16$ for each trial.
The gradient approximation from~\eqref{eq:linear-approx} is used. 
On each trial, we count the number of times the distortion function $g$ is called.
The results are shown in Figure~\ref{fig:cost} where it seen that the number
of calls actually tends to decrease as the allowed nonlinearity is increased,
indicating that the control landscape does not become more difficult to navigate.


In conclusion, we have presented an optimization framework that permits 
the design of robust quantum control sequences that account for general 
simulatable distortions by classical control hardware. 
We have demonstrated that even when distortions are nonlinear with respect to 
the input 
-- using the particular example of a nonlinear resonator circuit --
robust quantum control may still be achieved, and searching through the 
control landscape does not necessarily become more difficult.
Thus, classical control devices may be operated in their high power 
regime to permit fast high fidelity quantum operations, increasing the
number of gates that can be performed within the decoherence time of the 
quantum system.


\begin{acknowledgments}
Funding from Industry Canada, CERC, NSERC, Province of Ontario, CFI, and 
Mitacs is gratefully acknowledged.
I.H. thanks Hamid Mohebbi, Daniel Puzzuoli and Osama Moussa for helpful discussions.
\end{acknowledgments}


\bibliography{mainbib}


\newcommand{\complex}{\mathbb{C}}

\newcommand\T{\mathrm{T}}
\newcommand\rot{\mathrm{rot}}
\newcommand\eff{\mathrm{eff}}
\newcommand\init{\mathrm{init}}
\newcommand\target{\mathrm{target}}

\newcommand\ee{\mathrm{e}}
\newcommand\dd{\mathrm{d}}

\renewcommand{\Re}{\operatorname{Re}}
\renewcommand{\Im}{\operatorname{Im}}
\newcommand{\argmax}{\operatorname{arg\,max}}
\DeclarePairedDelimiter\ceil{\lceil}{\rceil}

\newcommand{\p}{\vec{p}}
\newcommand{\nrd}{n_{\text{rd}}}
\newcommand{\dtrd}{dt_{\text{rd}}}
\renewcommand{\prd}{\p_{\text{rd}}}

\renewcommand{\algorithmicrequire}{\textbf{Input:}}
\renewcommand{\algorithmicensure}{\textbf{Output:}}
\newcommand{\inlinecomment}[1]{\Comment {\footnotesize #1} \normalsize}
\newcommand{\linecomment}[1]{\State \(\triangleright\) {\footnotesize #1} \normalsize}

\onecolumngrid\appendix


\section{Examples of Common Distortions}

The distortion formalism outlined in the main body need not be applied to
only complicated systems; it is just as useful when applied to simple systems.
At a high level, having a general framework allows for problems to be tackled systematically,
allows for solutions found in one modality to be easily transferred to another,
and reduces development overhead as the system evolves.
In this section we outline some common, simple, but useful distortions, and 
show how they can be written down as discrete distortion operators.


\subsection{Composition}

First, it is worth noting the (perhaps obvious) fact that composing distortion
operators is easily implemented in this framework.
Suppose we have characterized the first half of our classical hardware 
with the discrete distortion operator 
$g_1:\real^N\otimes\real^K\rightarrow\real^{N'}\otimes\real^{K'}$ and the second
half with the operator $g_2:\real^{N'}\otimes\real^{K'}\rightarrow\real^M\otimes\real^L$.
We have been careful to make the domain of $g_2$ the same as the range of $g_1$.
Then the total distortion operator is given by the composition
\begin{align}
	g=g_2\circ g_1: \real^N\otimes\real^K & \rightarrow\real^M\otimes\real^L \\
	\vec{p} & \mapsto g_2(g_1(\vec{p})).
\end{align}

To find the Jacobian matrix of $g$ at point $\vec{p}$ we just need to use 
the multivariate chain rule,
\begin{equation}
	J_{\vec{p}}(g) = J_{g_1(\vec{p})}(g_2) \cdot J_{\vec{p}}(g_1),
\end{equation}
or in terms of indices,
\begin{equation}
	[J_{\vec{p}}(g)]_{m,l,n,k} = \sum_{n'=1}^{N'}\sum_{k'=1}^{K'}
		[J_{g_1(\vec{p})}(g_2)]_{m,l,n',k'}[J_{\vec{p}}(g_1)]_{n',k',n,k}.
\end{equation}


\subsection{Transfer Functions and Convolutions}

Linear electronic systems can be fully described by a 
\textit{transfer function} $\Phi(\omega)$.
This function gives a simple relationship between an input tone 
$X(\omega)$ at frequency $\omega$ and the resulting output tone 
$Y(\omega)$, namely $Y(\omega)=\Phi(\omega)X(\omega)$.
The magnitude of $H(\omega)$ represents the gain or attenuation, 
and the argument represents the phase shift.
Taking the inverse Fourier transform of this equation yields 
the convolution, $y(t)=(\phi\conv x)(t)$, where there may be factors 
of $2\pi$ missing due to convention.
The transfer function may be measured experimentally 
\cite{spindler_shaped_2012,gustavsson_improving_2013,rife_transfer-function_1989}, or may be
computed if a good model of the system is known.
In the main body, the formula for a discrete convolution operator 
arising from a time domain transfer function $\phi$ is shown.
Here, we derive it in slightly more detail.

To begin, the time domain version of the transfer function $\phi(t)$
results in the distortion operator $f$ defined as
\begin{equation}
	\beta(t)=f(\alpha)(t)=(\phi\conv\alpha)(t)=\int_{-\infty}^\infty \phi(t-\tau)\cdot\alpha(t) d\tau.
\end{equation}
Note that here $\phi(t)$ is a function whose values are $L\times K$ matrices.
In the usual context where $K=L$ and the $k'th$ output channel is mostly a distorted version
of the $k'th$ input channel, the the diagonals of $\phi(t)$ represent channel-wise distortions,
and the off-diagonals represent cross contamination between channels.
 
We can explicitly write the discretization and dediscretization operators mentioned in the 
main text as
\begin{align}
	f_1:~& \Lone{K} \to \real^M\otimes\real^K \nonumber \\
	& \quad \beta \mapsto(\beta(\delta t\cdot(1/2),...,\beta(\delta t\cdot(M-1/2))) \\
	f_2:~& \real^N\otimes\real^L \to \Lone{L} \nonumber \\
	& \quad (\vec{p}_1,...,\vec{p}_N) \mapsto \sum_{n=1}^N \vec{p}_n\cdot \operatorname{Top}(t-dt\cdot(n-1/2))
\end{align}
where $\operatorname{Top}$ is the $L$-dimensional top hat function,
\begin{equation}
    T(t) = \begin{cases} 
        (1,1,...,1) & 0\leq t < dt \\
        (0,0,...,0) & \text{else},
    \end{cases}
\end{equation}
and the factors of $1/2$ appear so that we are sampling the midpoint of each step.
We are also using the convention that an element of $\vec{p}\in\real^N\otimes\real^K$ is thought 
of as a vector $\vec{p}=(\vec{p}_1,...,\vec{p}_N)$ of vectors, where each $\vec{p}_k \in\real^K$.
This means our discretized distortion operator will be 	$g=f_1\circ f \circ f_2$.
Discretizing the input we get
\begin{align}
    f(f_2(\vec{p}))(t) & = \sum_{n = 1}^{N} \int_{(n-1) dt}^{n dt} \phi(t - \tau)\cdot \vec{p}_n\ \dd\tau \\
                            & \equiv \sum_{n = 1}^{N} \phi_n(t)\cdot\vec{p}_n,
\end{align}
which we then discretize the output of, to get
\begin{align}
    [(f_1 \circ f \circ f_2)(\vec{p})]_{m,l} & = \sum_{n=1}^{N} \sum_{k=1}^{K} [\phi_n((m-1/2) \delta t)]_{l,k}p_{n,k} \\
                                              & \equiv \sum_{n=1}^{N} \sum_{k=1}^{K} \phi_{m,l,n,k} p_{n,k},
\end{align}
for all $1\leq m\leq M$ and $1\leq l\leq L$ 
where 
\begin{align}
    \phi_{m,l,n,k} &= [\phi_n((m-1/2) \delta t)]_{l,k} \\
               &= \int_{(n-1) dt}^{n dt} [\phi((m-1/2) \delta t - \tau)]_{l,k}\ \dd\tau 
    \label{eq:convolution_mtx_element}
\end{align}
Letting $\tilde{\phi}\in\real^M\otimes\real^L\otimes\real^N\otimes\real^K$ 
be the tensor with entries $\phi_{m,l,n,k}$ gives 
\begin{align}
    g(\vec{p})=(f_1 \circ f \circ f_2)(\vec{p})=\tilde{\phi}\cdot\vec{p}
\end{align}
as a compact representation of the discretized distortion operator, where
the dot represents a contraction over the indices $n$ and $k$.

The elements of the Jacobian matrix $J(g)$ are now easily computed as
\begin{align}
    [J(g)]_{m,l,n,k} & = \frac{\partial (g(\vec{p}))_{m,l}}{\partial p_{n,k}} 
                        = \frac{\partial (\tilde{\phi}\cdot\vec{p})_{m,l}}{\partial p_{n,k}}
                        = \frac{\partial \sum_{n'=1}^N\sum_{k'=1}^K \phi_{m,l,n',k'}p_{n',k'}}{\partial p_{n,k}} 
                        = \phi_{m,l,n,k} \\
\intertext{so that}
    J(f_\phi) & = \tilde{\phi}.
\end{align}


\subsection{Finite Rise Times}
\label{sec:finite-rise-times}

A simple special case of the general convolution discussed in the previous subsection 
is a rise time acting independently on each control channel.
This will cause the rising edge of a square input pulse to be smoothed over
with an exponential of time constant $\tau$, and the trailing edge to decay back to zero 
with an exponential of the same time constant.
Such a transfer function arises, for example, from a simple $RL$ circuit, where the time constant
will be given by $\tau=L/R$.

\begin{center}
\begin{figure}
\includegraphics[scale=0.6]{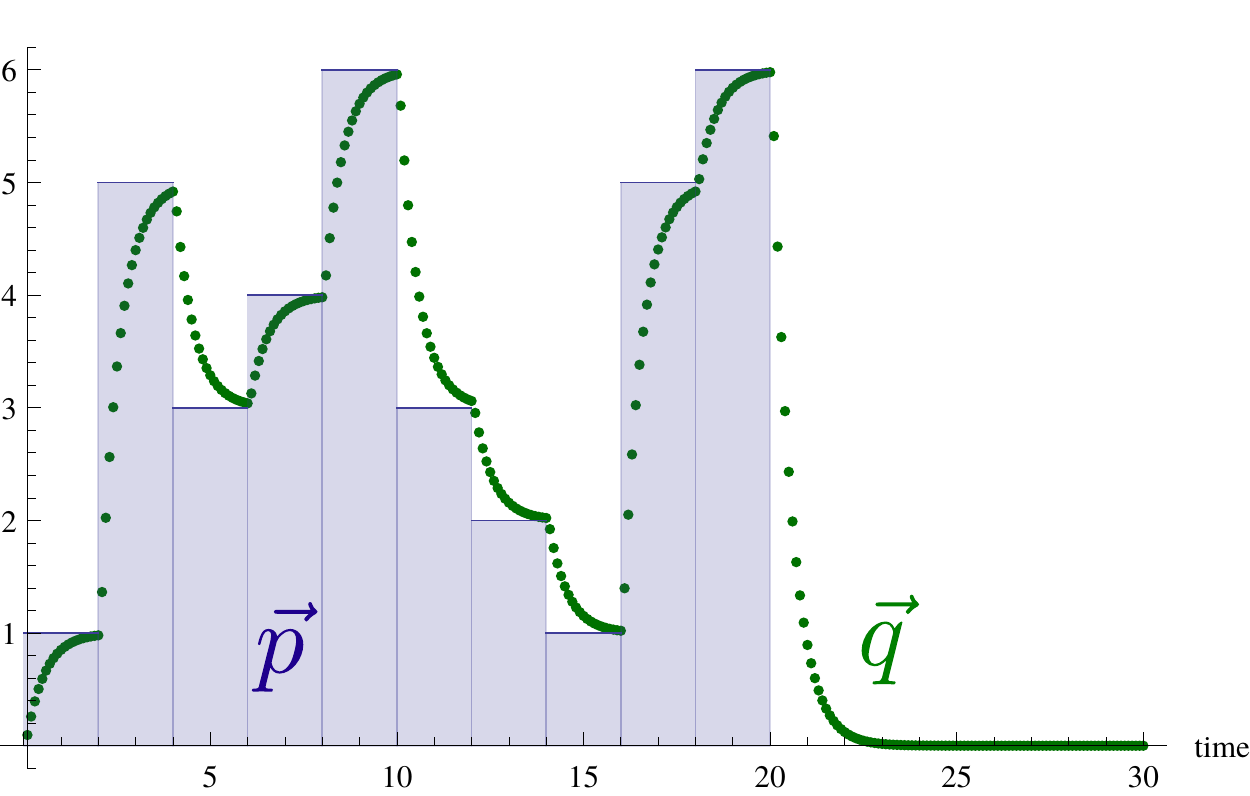}
\caption{
An example of the application of a discrete convolution distortion $g$ to
an input pulse with $N=10$ time steps. We have $dt=2$, and the output space has 20 time steps
per input time step, thus $\delta t=0.1$.
}
\label{fig:discrete_conv}
\end{figure}
\end{center}

Given a rise time $\tau_c^k$ acting independently on each of the $1\leq x\leq K=L$ control channels
gives the time-domain transfer function as
\begin{equation}
	\phi_{l,k}(t)=
	\begin{cases}
		\frac{1}{\tau_c^k}e^{-t/\tau_c^k} & l=k \text{ and } t\geq 0 \\
		0 & \text{else}
	\end{cases}
\end{equation}
which, when the integral from \eq{convolution_mtx_element} is performed and simplified 
(by Mathematica in this case), 
results in the discrete convolution
\begin{equation}
	[\tilde{\phi}]_{m,l,n,k}=
    \begin{cases}
       \delta_{l,k}(e^{dt/\tau_c^k}-1)e^{\frac{t_{n-1}-t'_m}{\tau_c^k}} 
       	   & t_n < t'_m \\
       \delta_{l,k}\left(1-e^{\frac{t_{n-1}-t'_m}{\tau_c^k}}\right)
           & (t_n=t'_m) \vee ((t_n>t'_m)  \wedge (n=1 \vee t_{n-1}<t'_m)) \\
       0 & \text{else}
    \end{cases}
\end{equation}
for each $1\leq k\leq K=L$, where $t_n=n dt$ and $t'_m=(m-1/2) \delta t$.
This is illustrated in \fig{discrete_conv} with $K=L=1$.

\subsubsection{Example of Robustness to Rise Time}

We consider designing a CNOT gate for two qubits with an internal Hamiltonian
\begin{equation}
	H =
	\frac{\omega_1}{2}\sigma_z^1 + \frac{\omega_2}{2}\sigma_z^2
	+ \frac{J}{4} (\sigma_x^1\sigma_x^2 + \sigma_y^1\sigma_y^2 + \sigma_z^1\sigma_z^2)
\end{equation}
and control Hamiltonians
\begin{equation}
	\left\{ 
		H_\text{x} = \sigma_x^1 + \sigma_x^2,~~
		H_\text{y} = \sigma_y^1 + \sigma_y^2
	\right\}
\end{equation}
where $\omega_1=-2\pi\cdot 15$, $\omega_2=+2\pi\cdot 15$, $J=2\pi\cdot 50$, 
and the amplitudes of the control Hamiltonian are bounded by $2\pi\cdot50$.
This is the style of Hamiltonian found in liquid state NMR homonuclear samples
\cite{levitt_spin_2001}.
We use $N=30$ input time steps of length $dt=0.005$, and $M=2N+\ceil{10 \tau/dt}$ 
output time steps of length $dt/2$.
Here, $\tau$ is the characteristic exponential rise time of both control channels,
as defined in \sec{finite-rise-times}.

\begin{center}
\begin{figure}
\includegraphics[width=\textwidth]{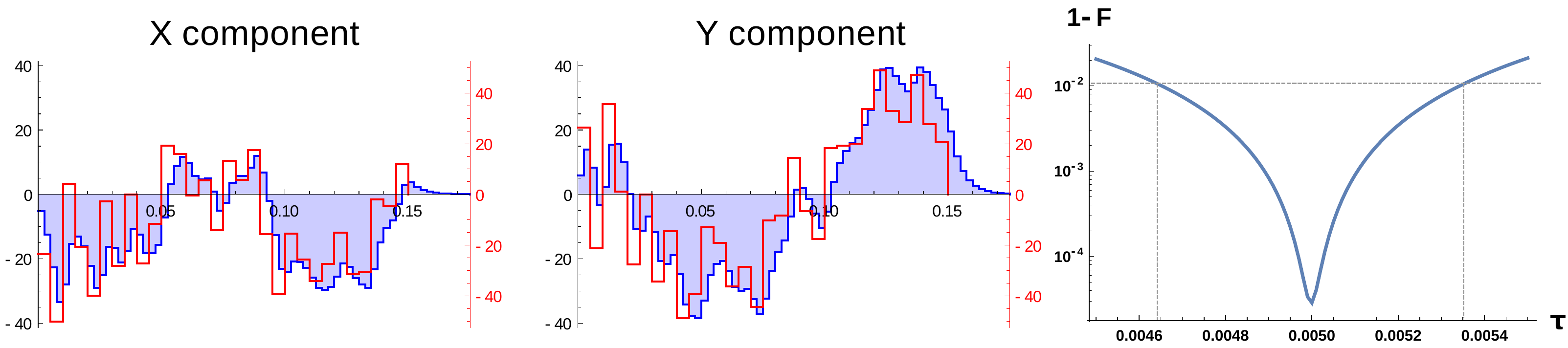}
\caption{
The pulse envelope of a CNOT gate at an exponential rise time value of $\tau=0.005$.
The robustness curve, in terms of one minus average fidelity, is shown to the 
right as a function of $\tau$.
}
\label{fig:robust-ringdown}
\end{figure}
\end{center}

To make the resulting pulse sequence robust against the value $\tau$, we set
the objective function of the optimization problem to be a convex combination
of objective functions, each with a different value of $\tau$, as explained
in \sec{static-param-dist}.
The results are shown in \fig{robust-ringdown}, where it is seen that
$F>0.99$ is achieved in a region about $\pm 7\%$ around a nominal value of
$\tau=0.005$.

\subsection{Crosstalk}
\label{sec:crosstalk}

Crosstalk is the phenomenon where a signal sent along one control channel is
overheard by other control channels.
As alluded to earlier, this can be fully accounted for (in the case of linear
controllers) by the off-diagonal elements of the transfer function $\phi$.
This may be overkill as crosstalk can often be accurately modelled
as one control line seeping into each of the other control lines with 
attenuation factors that are constant in time.
See Reference~\cite{barends_superconducting_2014} for example, where 
crosstalk between five coupled superconducting qubits is observed.

To model this situation, we consider that our quantum system has $I$ subsystems (or qubits)
each with $L=K$ control channels.
The Hamiltonian is given by
\begin{equation}
H=H_\text{int}+\sum_{i=1}^I\sum_{l=1}^L q_{i,l}(t)H_{i,l}
\end{equation}
where $H_\text{int}$ is the internal Hamiltonian, containing all coupling terms,
and $q_{i,l}(t)$ and $H_{i,l}$ are the $l^\text{th}$ control envelope and 
Hamiltonian of the $i^\text{th}$ system.
There are now a total of $I\cdot L$ controls, and so instead of indexing the control
indeces by single numbers, $k$ and $l$, we index them by tuples, $(i,k)$ and $(i,l)$.
Since this distortion is independent of time, we set $\delta t=dt$ and $M=N$.
Ideally, we would have $p_{n,(i,k)}=q_{n,(i,k)}$ representing the fact that the 
$(i,k)^\text{th}$ control signal is sent exactly to the $(i,k)^\text{th}$ Hamiltonian
at each time step $n$.
With crosstalk included, the $(i,k)^\text{th}$ Hamiltonian actually sees a linear combination
of each of every control line,
\begin{equation}
q_{n,(i,l)}=\sum_{j=1}^I\sum_{k=1}^K \chi_{(i,l),(j,k)}p_{n,(j,k)},
\end{equation}
where $\chi_{(i,l),(j,k)}$ is the fraction of the $(j,k)^\text{th}$ control line seen on 
the $(i,l)^\text{th}$ control.
More compactly,
\begin{equation}
\vec{q}=g(\vec{p})=\chi\cdot\vec{p}
\label{eq:crosstalk-distortion}
\end{equation}
where the dot in this case represents contraction over the indices $j$ and $k$.

The ideal $\chi$ tensor is $\chi_{(i,l),(j,k)}=\delta_{i,j}\delta_{l,k}$.
As an example, if there are $4$ qubits each with two controls, x and y, then 
in matrix format the ideal tensor reads
\begin{equation}
	\chi_\text{ideal} = 
	\begin{blockarray}{cccccccccc}
		~&~~~~& \BAmulticolumn{2}{c}{Q1} & \BAmulticolumn{2}{c}{Q2} 
		      & \BAmulticolumn{2}{c}{Q3} & \BAmulticolumn{2}{c}{Q4} \\
		&& x & y & x & y & x & y & x & y \\	
		\begin{block}{cc(cc|cc|cc|cc)}
			\multirow{2}{*}{Q1} 
			& x & 1 & 0 & 0 & 0 & 0 & 0 & 0 & 0 \\
			& y & 0 & 1 & 0 & 0 & 0 & 0 & 0 & 0 \\
			\BAhhline{&&--------} 
			\multirow{2}{*}{Q2} 
			& x & 0 & 0 & 1 & 0 & 0 & 0 & 0 & 0 \\
			& y & 0 & 0 & 0 & 1 & 0 & 0 & 0 & 0 \\
			\BAhhline{&&--------} 
			\multirow{2}{*}{Q3} 
			& x & 0 & 0 & 0 & 0 & 1 & 0 & 0 & 0 \\
			& y & 0 & 0 & 0 & 0 & 0 & 1 & 0 & 0 \\
			\BAhhline{&&--------} 
			\multirow{2}{*}{Q4} 
			& x & 0 & 0 & 0 & 0 & 0 & 0 & 1 & 0 \\
			& y & 0 & 0 & 0 & 0 & 0 & 0 & 0 & 1 \\
		\end{block}
	\end{blockarray}
\end{equation}
If we add a crosstalk term between adjacent qubits, where an $x$ control only
talks to adjacent $x$ controls and similar for $y$ controls, the tensor might
look like
\begin{equation}
	\chi_\text{nearest neighbour} = 
	\begin{blockarray}{cccccccccc}
		~&~~~~& \BAmulticolumn{2}{c}{Q1} & \BAmulticolumn{2}{c}{Q2} 
		      & \BAmulticolumn{2}{c}{Q3} & \BAmulticolumn{2}{c}{Q4} \\
		&& x & y & x & y & x & y & x & y \\	
		\begin{block}{cc(cc|cc|cc|cc)}
			\multirow{2}{*}{Q1} 
			& x & 1 & 0 & 0.2 & 0 & 0 & 0 & 0 & 0 \\
			& y & 0 & 1 & 0 & 0.3 & 0 & 0 & 0 & 0 \\
			\BAhhline{&&--------} 
			\multirow{2}{*}{Q2} 
			& x & -0.1 & 0 & 1 & 0 & 0.5 & 0 & 0 & 0 \\
			& y & 0 & 0.15 & 0 & 1 & 0 & 0.4 & 0 & 0 \\
			\BAhhline{&&--------} 
			\multirow{2}{*}{Q3} 
			& x & 0 & 0 & -0.2 & 0 & 1 & 0 & 0.2 & 0 \\
			& y & 0 & 0 & 0 & -0.2 & 0 & 1 & 0 & 0.3 \\
			\BAhhline{&&--------} 
			\multirow{2}{*}{Q4} 
			& x & 0 & 0 & 0 & 0 & 0.23 & 0 & 1 & 0 \\
			& y & 0 & 0 & 0 & 0 & 0 & 0.7 & 0 & 1 \\
		\end{block}
	\end{blockarray}
\end{equation}

As is clear from Equation~\ref{eq:crosstalk-distortion}, the Jacobian of 
this distortion operator is simply given by
\begin{equation}
J_{\vec{p}}(g)=\chi.
\end{equation}
A pulse could be designed to be robust against errors in the crosstalk tensor 
by including a distribution over crosstalk tensors, or perhaps just a 
distribution over some of its values, dependently or independently, using
the method described in \sec{static-param-dist}.


\subsubsection{Crosstalk Example}

We design a $\pi/2$ gate about $x$ on the third of four qubits arranged in a line,
with the other three qubits performing the identity operation.
Each qubit has an $x$ and $y$ control, $\{H_{i,x}=\sigma_x^i,H_{i,y}=\sigma_y^i\}$,
and the internal Hamiltonian is given by
\begin{equation}
	H =
	\sum_{|i-j|=1}\omega_{ij}\sigma_z^i\sigma_z^j
\end{equation}
where $\omega_{ij}=2\pi\cdot 20$MHz and the input control amplitudes are limited to
$2\pi\cdot 40$MHz.
We use a crosstalk tensor
\begin{equation}
	\chi = 
	\begin{blockarray}{cccccccccc}
		~&~~~~& \BAmulticolumn{2}{c}{Q1} & \BAmulticolumn{2}{c}{Q2} 
		      & \BAmulticolumn{2}{c}{Q3} & \BAmulticolumn{2}{c}{Q4} \\
		&& x & y & x & y & x & y & x & y \\	
		\begin{block}{cc(cc|cc|cc|cc)}
			\multirow{2}{*}{Q1} 
			& x &  1 & 0 & 0.3 & 0.001 & 0.05 & 0 & 0.001 & 0 \\
			& y & 0 & 1 & 0 & 0.1 & 0 & 0.01 & 0 & 0.001 \\
			\BAhhline{&&--------} 
			\multirow{2}{*}{Q2} 
			& x & 0.25 & 0 & 1 & 0 & 0.3 & -0.005 & 0.04 & 0 \\
			& y & 0 & 0.2 & 0 & 1 & 0 & 0.4 & 0 & 0 \\
			\BAhhline{&&--------} 
			\multirow{2}{*}{Q3} 
			& x & 0 & 0 & 0.2 & 0 & 1 & 0 & -0.2 & 0 \\
			& y & 0 & -0.04 & 0 & 0.2 & 0 & 1 & 0 & 0.3 \\
			\BAhhline{&&--------} 
			\multirow{2}{*}{Q4} 
			& x & 0.001 & 0 & 0.04 & 0 & 0.3 & 0 & 1 & 0 \\
			& y & 0 & 0 & 0 & 0.07 & 0 & -0.3 & 0 & 1 \\
		\end{block}
	\end{blockarray}
\end{equation}

\begin{center}
\begin{figure}
\includegraphics[width=\textwidth]{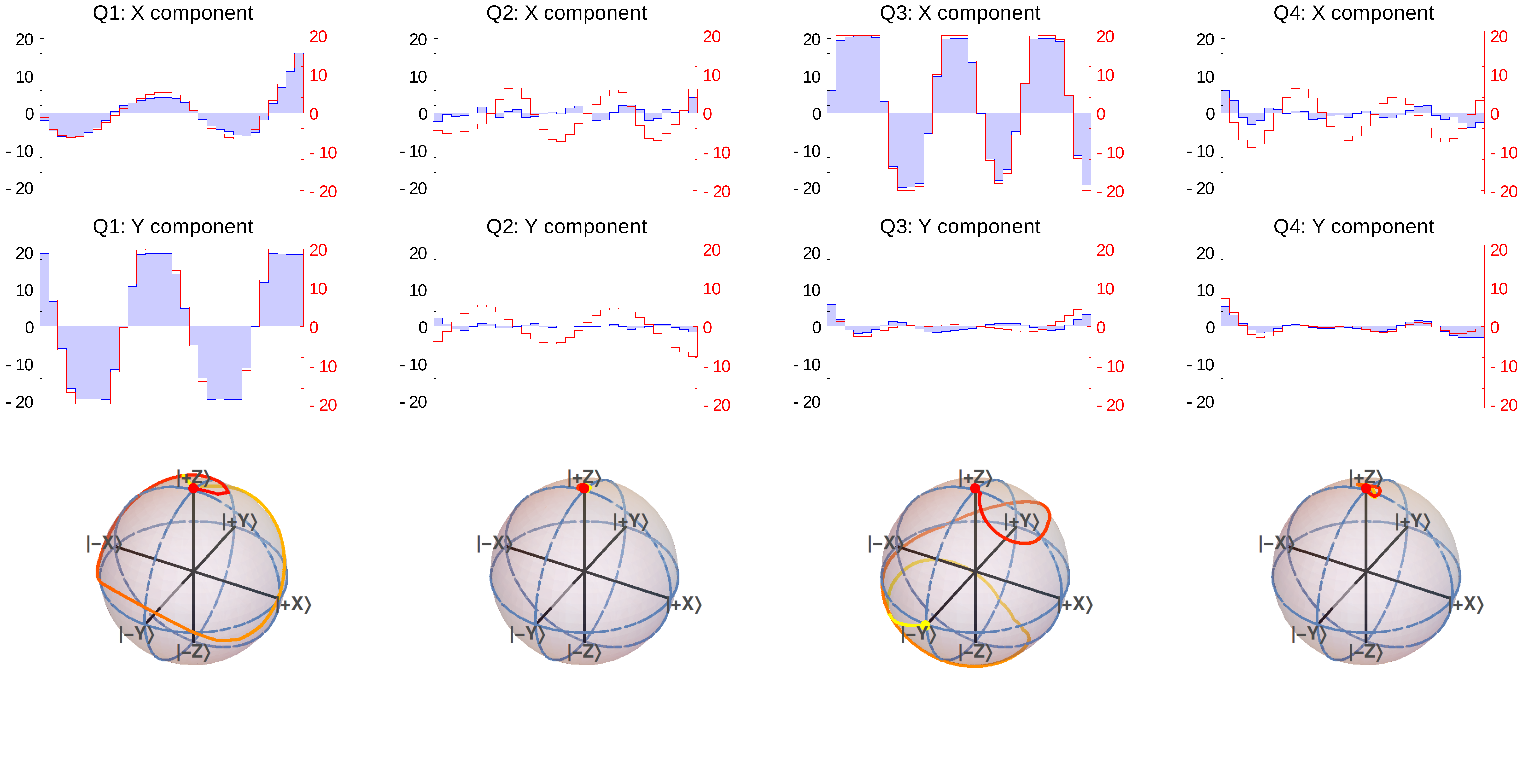}
\caption{
(color online)
The pulse envelopes and Bloch sphere trajectories of a $\pi/2)_x$ gate on the third qubit.
The (unfilled) red curves represent the input pulse, and the (filled) blue curves represent the output pulse
seen by the quantum system.
}
\label{fig:crosstalk}
\end{figure}
\end{center}

Using this crosstalk distortion tensor, a pulse with average fidelity 
$F=0.9999$ was found and is shown in \fig{crosstalk}.
 
\section{The Rotating Frame of the Circuit}
\label{sec:rot-circuit}

\begin{figure}
  \centering
  \includegraphics[width=1.5in]{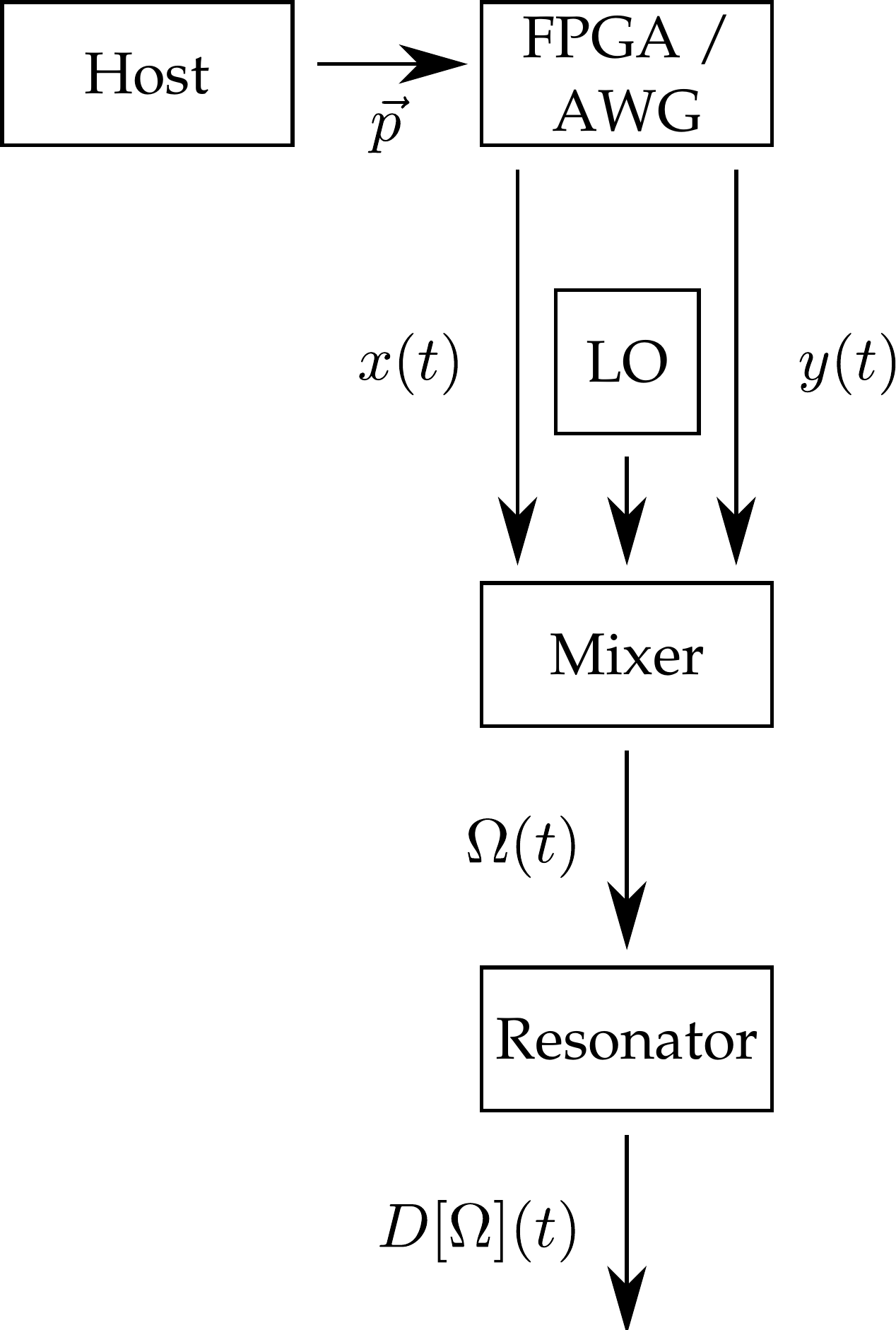}
  \caption{\label{fig:mixing} Configuration of the microwave mixing components in relation to pulse distortion operators.}
\end{figure}

A spin in a large static magnetic field $\gamma B_0=\omega_0$ with a transverse time dependent field $\gamma B_1(t)=2\Omega(t)$ 
will evolve under the Hamiltonian
\begin{equation}
    H=\frac{\omega_0}{2}\sigma_z+2\frac{\Omega(t)}{2}\sigma_x.
\end{equation}
Since $\omega_0$ is taken to be the dominant term, in analogy to a wide range of experimental settings,
it is helpful to enter the rotating frame generated by $H_\rot \defeq \omega_r \sigma_z / 2$,
where $\omega_r \defeq [\omega_0 + \delta\omega]$.
In doing so, we will suppose that
\begin{align*}
\Omega(t) & = \omega_1(t) \cos(\omega_r t + \phi(t)) \\
          & = \omega_1(t) \cdot \frac{\ee^{\ii t (\omega_r t + \phi(t))} + \ee^{-\ii t (\omega_r t + \phi(t))}}{2},
\end{align*}
representing that $\Omega$ is produced by mixing a modulating signal
with an oscillator at $\omega_0 + \delta\omega$ (see Figure \ref{fig:mixing}).
We will later relate this model to the in-phase and quadrature control fields.

In the frame of $H_\rot$, the effective Hamiltonian $H_\eff$ is given by
\begin{align*}
    H_\eff(t) & = \ee^{+\ii H_\rot t} H(t) \ee^{-\ii H_\rot t} - H_\rot \\
              & = \Omega(t) \ee^{+\ii H_\rot t} \sigma_x \ee^{-\ii H_\rot t} - \delta\omega\,\sigma_z  \\              
              & = \Omega(t) [\cos(\omega_r t) \sigma_x - \sin(\omega_r t) \sigma_y] - \delta\omega\,\sigma_z. \\
\intertext{Discarding the terms which oscillate at $2 \omega_r$ (that is, the rotating wave approximation), we can rewrite this
in terms of $\omega_1$ and $\phi$ instead,
}
    H_\eff(t) & = \frac{\omega_1(t)}{2} [\cos(\phi(t)) \sigma_x + \sin(\phi(t)) \sigma_y] - \delta\omega\,\sigma_z.
\end{align*}
Since $\Omega$ is often produced by mixing, as noted above, we can also represent the
rotating frame control using two real control fields $\omega_x(t)$ and $\omega_y(t)$,
\begin{equation}
  \omega_x(t) = \Re[\omega_1(t) \ee^{\ii \phi(t)}] \qquad
  \omega_y(t) = \Im[\omega_1(t) \ee^{\ii \phi(t)}].
  \label{eq:rabi-fields-pre}
\end{equation}
Using these control fields, we match the definition of $H_\eff$ in the main body, given
as Equation~11.

This rotating frame can be analogously extended into the circuit dynamics as well.
The nutation frequency of the spins is proportional to the magnetic field generated by the 
inductor via the gyromagnetic ratio, which is in turn proportional to the current passing 
through the inductor, thus
\begin{equation}
    \Omega(t) = \kappa I_L(t),
\end{equation}
where the exact value of $\kappa$ depends on the relevant geometry.
This along with Equation~\ref{eq:rabi-fields-pre} leads us to the relationships
\begin{align}
	\omega_x(t) 
		= \Re[\omega_1(t) \ee^{\ii \phi(t)}]
		= \Re[\Omega(t) \ee^{-i\omega_r t}]
		= \kappa \Re[I_L \ee^{-i\omega_r t}]
		= \kappa \Re[\tilde{I}_L] \nonumber \\
	\omega_y(t) 
		= \Im[\omega_1(t) \ee^{\ii \phi(t)}]
		= \Im[\Omega(t) \ee^{-i\omega_r t}]
		= \kappa \Im[I_L \ee^{-i\omega_r t}]
		= \kappa \Im[\tilde{I}_L],
	\label{eq:rabi-fields}
\end{align}
where  $\tilde{I}_L(t) = \ee^{-\ii\omega_1 t} I_L(t)$ and we have ignored pieces rotating at $2\omega_r$ in the calculation.
We recover the expression in the main body, such that by solving the differential equation
$\dot{\vec{x}}(t) = A(\vec{x})\vec{x} + \tilde{V}_s(t)\vec{b}$
for a complex driving function $\tilde{V}_s(t)$, we can find the
rotating-frame unitary action $U(t) = \T\exp(-\ii\int_0^t H_\eff(t))$ of that pulse.


\section{The Discretized Distortion Operator Due to a Resonator Circuit}


\subsection{Definition of the Distortion Operator}

We define the distortion operator $g:\real^N\otimes\real^2\rightarrow\real^M\otimes\real^2$
corresponding to the non-linear resonator circuit shown in the main body.
Note that the input to $g$ will have units of volts, and the output of $g$ will have units of Hz.
We use a uniform input discretization time $dt$ and a uniform output discretization time $\delta t$.
Given an input pulse $\vec{p}\in\real^N\otimes\real^2$, we define the complex vector $\tilde{p}\in\complex^N$
by $\tilde{p}_{n}=p_{n,1}+ip_{n,2}$.
Setting $n(t)=\ceil*{\frac{t}{dt}}$, we define
\begin{equation}
\label{eq:smooth_forcing}
\alpha(t) = \tilde{p}_{n(t)}
\end{equation}
in the case where we add no rise time to the forcing term, or
\begin{equation}
\label{eq:smooth_forcing}
\alpha(t) = \tilde{p}_{n(t)-1} + (\tilde{p}_{n(t)}-\tilde{p}_{n(t)-1})(1-e^{-\frac{t-n(t)dt}{\tau_r}}).
\end{equation}
in the case where we include a finite rise time on the forcing with timescale $\tau_r$.
Note that we must have $\tau_r \ll dt$ for the function $\alpha$ to be (approximately) continuous.
This limitation could easily be overcome with a more sophisticated definition of $\alpha$, for example,
by using a convolution operator.
Also note we are using the convention $\tilde{p_i}=0$ for $i<1$ or $i>N$.

Now we solve the vector differential equation
\begin{equation}
	\dot{x}=A(x)x+\alpha(t) b,
	\label{eq:resonator_de}
\end{equation}
where
\begin{equation}
	x=
	\begin{bmatrix}
		\tilde{I}_L \\ \tilde{V}_{C_m} \\ \tilde{V}_{C_t}
	\end{bmatrix}
	\quad\quad\quad
	A(x)=
	\begin{bmatrix}
		-\frac{R}{L} & 0 & \frac{1}{L} \\
		0 & \frac{-1}{R_L C_m} & \frac{1}{R_L C_m} \\
		\frac{-1}{C_t} & \frac{-1}{R_L C_t} & \frac{1}{R_L C_t}
	\end{bmatrix}
	- i\omega_r \I
	\quad\quad\quad
	b=
	\begin{bmatrix}
		0 \\ \frac{1}{R_L C_m} \\ \frac{1}{R_L C_t}
	\end{bmatrix}
\end{equation}
with Mathematica 10's function NDSolve.
By default, this function dynamically chooses the step size and 
switches between solvers, of both the implicit and explicit 
time stepping variety, to ensure that the solution is accurate and
stable.
An interpolating function for $\tilde{I}_L(t)$ is returned.
Recalling Equation~\ref{eq:rabi-fields}, we sample the real 
and imaginary parts of $\tilde{I}_L(t)$
at a rate $\delta t$ to obtain $\vec{q}$:
\begin{align}
q_{m,1}=\kappa \Re \tilde{I}_L(\delta t(m-1/2)) \nonumber \\
q_{m,2}=\kappa \Im \tilde{I}_L(\delta t(m-1/2))
\end{align}


\subsection{Jacobian of the Non-linear Resonator Distortion}

To populate elements of the Jacobian tensor $J_{\vec{p}}(g)$, we are interested 
in approximating partial derivatives of the form
\begin{equation}
	\frac{\partial g_{m,l}}{\partial p_{n,k}}
\end{equation}
where $g$ is the distortion corresponding to the non-linear resonator circuit.
The most straight forward way of approximating such would be to use a central 
difference formula
\begin{equation}
	\frac{\partial g_{m,l}}{\partial p_{n,k}}
	\approx
	\left[
		\frac{g(\vec{p}+\epsilon \vec{e}_{n,k})-g(\vec{p}-\epsilon\vec{e}_{n,k})}{2\epsilon}
	\right]_{m,l},
\end{equation}
where $\vec{e}_{n,k}$ is the unit vector in the $\{n,k\}$ direction, and 
$\epsilon>0$ is a small number that is greater than the precision of the DE solver.
Such an approximation would require $2NK$ calls to the DE solver.
It is also numerically unstable as it involves the difference of two numerical 
DE solutions whose forcing terms are only slightly different; $\epsilon$ would
have to be very carefully tuned and may have no reliable value at all, especially
when searching for high fidelity pulses.

If we consider the approximation 
$g(\vec{p}\pm\epsilon\vec{e}_{n,k})\approx g(\vec{p})\pm g(\epsilon\vec{e}_{n,k})$
the central difference reduces to
\begin{equation}
	\frac{\partial g_{m,l}}{\partial p_{n,k}}
	\approx
	\left[
		g(\epsilon \vec{e}_{n,k})/\epsilon
	\right]_{m,l}.
	\label{eq:linear-approx}
\end{equation}
which is the approximation quoted in the main body.
Importantly, this approximation does not depend on the current pulse $\vec{p}$ 
and can therefore be precomputed eliminating the $2NK$ calls to $g$ (i.e. DE 
solver calls) per ascension step.

An exact method to compute these partial derivatives is derived below, which
will take $N*K+1$ calls to the DE solver to compute the entire Jacobian matrix.
Begin with the resonator differential equation (Equation~\ref{eq:resonator_de})
\begin{equation}
	\dot{x}=A(x)x+\alpha(t) b.
\end{equation}
As discussed, we have
\begin{align}
	[g(\vec{p})]_{m,1} = \kappa\re \tilde{I}_L(t_m) \equiv h_1(x(t_m)) \nonumber \\
	[g(\vec{p})]_{m,2} = \kappa\im \tilde{I}_L(t_m) \equiv h_2(x(t_m))
\label{eq:flux-formula}
\end{align}
where
$t_m=(m-1/2)\delta t$.
Thus it is clear that the difficult part of computing
$\frac{\partial g_{m,l}}{\partial p_{n,k}}$ is computing 
$\frac{\partial \tilde{I}_L}{\partial p_{n,k}}$, or more generally 
$\frac{\partial x}{\partial p_{n,k}}$.

We derive a set of $K*N=2N$ secondary partial differential 
vector equations whose time sampled solutions produce the necessary partial derivatives.
To do this we just take the partial derivative $\frac{\partial}{\partial p_{n,k}}$
of Equation~\ref{eq:resonator_de}, which gives as the $l^\text{th}$ component of 
the $(n,k)^\text{th}$ equation
\begin{align}
	\frac{\partial}{\partial p_{n,k}}\frac{\partial x_l}{\partial t}
		= \frac{\partial A_{l,l'}}{\partial x_{l''}} \frac{\partial x_{l''}}{\partial p_{n,k}}x_{l'} 
		+ [A(x)]_{l,l'}\frac{\partial x_{l'}}{\partial p_{n,k}} + T_{n,k}b_l
\label{eq:deriv1}
\end{align}
where Einstein summation notation is used and (in the case $\tau_r=0$)
\begin{equation}
	T_{n,k}(t)=\begin{cases}
		0 & 0\leq t \leq dt \\
		\vdots & \\
		\delta_{1,k}+i\delta_{2,k} & (n-1)dt\leq t \leq ndt \\
		\vdots & \\
		0 \leq t \leq Ndt
	\end{cases}.
\end{equation}
Denote 
\begin{align}
	y_{n,k}(t)=\frac{\partial x}{\partial p_{n,k}}(t) \nonumber \\
	[A'(x)]_{l,l''} = \frac{\partial A_{l,l'}}{\partial x_{l''}} x_{l'}
\end{align}
 and commuting the partial derivatives,
the components Equation~\ref{eq:deriv1} can be rewritten as the non-linear vector PDE
\begin{equation}
	\dot{y}_{n,k} = [A'(x)+A(x)] y_{n,k} + T_{n,k}(t)b
\label{eq:deriv-de}
\end{equation}
where $x(t)$ is the solution to Equation~\ref{eq:resonator_de}.
Therefore once $x(t)$ has been computed, we can plug it into each of the DEs for $y_{n,k}$,
solve them with the initial condition $y_{n,k}((n-1)dt)=0$ (by causality 
$y_{n,k}=0$ for $t<(n-1)dt$) and we arrive at the exact formula
\begin{align}
	\frac{\partial g_{m,l}}{\partial p_{n,k}} 
		= \frac{\partial h_l(x(t))}{\partial p_{n,k}} \bigg|_{t=t_m} \nonumber \\
		= \frac{\partial h_l}{\partial x_{l'}}\frac{\partial{x_{l'}}}{\partial p_{n,k}} \bigg|_{t=t_m} \nonumber \\
		= \frac{\partial h_l}{\partial x_{l'}} [y_{n,k}(t_m)]_{l'}
\end{align}
where $h_l$ was defined implicitly in Equation~\ref{eq:flux-formula} and each 
$\frac{\partial h_l}{\partial x_{l'}}$ is easy to compute.

If we take the Taylor series of $A(x)$ about $x=0$, we have
\begin{equation}
A(x)=A_0 + A_1(x) + A_2(x) + \hdots
\end{equation}
where each $A_p$ is a matrix polynomial in the coordinates of $x$ with all terms 
having order exactly $p$.
The $0^\text{th}$ order approximation of Equation~\ref{eq:deriv-de} gives
\begin{equation}
	\dot{y}_{n,k} = A_0 y_{n,k} + T_{n,k}(t)b.
\end{equation}
In this form we see that $y_{n,k}$ is just the same as $x$ where the DE for $x$,
Equation~\ref{eq:deriv-de}, has been linearized and the forcing is the top hat $T_{n,k}$:
$y_{n,k}=x|_{A=A_0,\alpha=T_{n,k}}$.
The linearization condition $A=A_0$ is approximately the same as the guarantee 
$\|A(x)-A_0\| \ll 1$, which can be met by setting $\alpha=\epsilon T_{n,k}$ with
$\epsilon$ chosen so that $\left\|A(\frac{\epsilon\|b\|}{\|A_0\|})-A_0\right\| \ll 1$.
Therefore the zeroth order approximation to the Jacobian is
\begin{equation}
\frac{\partial g_{m,l}}{p_{n,k}} \approx \frac{g(\epsilon e_{n,k})}{\epsilon}.
\end{equation}
which is a somewhat more satisfying derivation of Equation~\ref{eq:linear-approx}.
 

\section{Ringdown Compensation}

A resonator or cavity with a large quality factor $Q$ will store
energy for times that are long compared to the time steps
that are used in pulse design. If this effect is not included in
optimization by integrating the distortion differential equation
for a sufficient period, then the integrated action of the pulse
on the quantum system will not be accurate. This can be dealt with
by defining the image of the distortion operator to represent a
longer time interval than the domain, but this is inconvenient
in experimental practice, where we would like to turn off a
pulse quickly. Thus, a better alternative is to actively compensate
for the ringdown introduced by large $Q$, and to demand that the
distorted pulse goes to zero at a given time step.

For a resonator with only
linear elements, this problem has been solved \cite{borneman_bandwidth-limited_2012} by appealing to
the transfer function
$h : \real^M \to \real^K$,
\begin{equation}
  g[\vec{p}] = f_1 [f_2(\vec{p}) \star h]
\end{equation}
where $\star$ is the convolution operator. For the case $M = K = 1$,
the transfer function takes on the simple form
\begin{equation}
  h(t) = A \ee^{-t / \tau_c}
\end{equation}
for some amplitude $A$ and where $\tau_c = Q / \omega_0$
is a time constant. In this case, it is easy to append an
additional pulse segment of amplitude
\begin{equation}
p_{K+1} = - A \frac{g[\vec{p}]_m}{e^{\delta t / \tau_c} - 1},
\end{equation}
where $m$ is a time step index such that $t_m = t_K$.

In the nonlinear case, $Q$, $\omega_0$ and $A$ are not constant,
but depend on $\vec{p}$, and so more attention is required.
One solution is to modify the performance
functional to include the demand that the ringdown go to zero
by defining
\begin{equation}
  \Phi'_g(\vec{p}) := \Phi_g(\vec{p}) - \Omega_g(\vec{p}) = (\Phi - \Omega) \circ g.
\end{equation}
For ringdown compensation,
\begin{equation}
  \Omega_{\text{RD}} := \sum_{m = m_0}^{M} |p_m|^2,
\end{equation}
where $m_0$ is the time step index at which we start demanding
that the solution goes to zero. The derivatives of this function
are easily found, such that $\vec{\nabla}\Phi'$ is easy to compute
given $\vec{\nabla}\Phi$ and $J(g)$. Since a solution that both
has high fidelity with a unitary target and admits ringdown
compensation can be hard to find, we use the ringdown-compensation
method found in the next section to generate \emph{initial guesses}
which result in a small penalty $\Phi'_g(\vec{p})$.

Another solution is to include ringdown suppression in the distortion operator $g$ itself.
That is, given an input pulse $\vec{p}$, the forcing term $\alpha$ now includes not only
steps taken directly from $\vec{p}$, but also additional steps which are chosen
(according to the results from the next section) to eliminate the energy from the
cavity in a short period of time.
This was the method employed for the results shown in the main body of this Letter.

\subsection{Eliminating energy from a non-linear resonator}

Here, we derive a scheme to calculate the values of compensation steps to append to a pulse which act to remove the energy from a resonator on a timescale shorter than the ringdown time.

Write the equation of the circuit as 
\begin{equation}
\label{eq:circuit_de}
\dot{x}=Ax+\alpha b
\end{equation}
where $x$ is a vector of state variables for the circuit, $A$ is a matrix describing the circuit without forcing, $b$ is the forcing direction of the circuit, and $\alpha$ is a controllable scalar which sets the magnitude of the forcing.
We assume that we have already entered the frame rotating at the resonance frequency so that all quantities are complex, where real quantities correspond to in-phase components, and imaginary quantities correspond to quadrature components.
Note that for a non-linear circuit, $A$ will depend on the state of the system, that is, $A=A(x)$.
Moreover, $\alpha$ can be time dependent, $\alpha=\alpha(t)$.

Our goal is as follows: start with an undistorted pulse $\p_0$ and append $\nrd$ steps of length $\dtrd$ to form the undistorted pulse $\p=[\p_0, \prd]$ which cause the distorted pulse $g(\p)$ to have near zero amplitude at the end of the last time step.
To simplify our task, we make the approximation that $A$ remains constant during each of the compensation steps, taking on a value corresponding to the state $x$ at the end of the previous time step.

The general solution to ~\ref{eq:circuit_de} is given by
\begin{equation}
x(t)=e^{tA}x_0 + \int_{0}^{t} \alpha(s)e^{(t-s)A}b~ds.
\end{equation}
Substituting our continuous forcing solution from equation~\ref{eq:smooth_forcing} and translating the time coordinate so that $t=0$ corresponds to the transition from the $(n-1)^\text{th}$ to the $n^\text{th}$ gives the solution
\begin{align}
x(t) &=
	e^{tA}x_0 
	+ e^{tA} \left[ \int_{0}^{t} e^{-sA}
		\left( \tilde{p}_{n-1} + (\tilde{p}_{n}-\tilde{p}_{n-1})(1-e^{-s/\tau_r}) \right)
		ds \right] b \nonumber \\
&=
	e^{tA}x_0 
	+ \left[
		\tilde{p}_n A^{-1}(e^{tA}-\I)
		- (\tilde{p}_n-\tilde{p}_{n-1})
			\left( A+\I/\tau_r \right)^{-1}
			\left( e^{tA}-e^{-t/\tau_r}\I \right)
	\right] b
\end{align}
in the region $t\in[0,\dtrd]$.
We wish to drive the state of the system, $x$, to 0.
Therefore, let's try to demand that at time $t=\dtrd$, $x$ becomes some fraction of its value at the end of the $(n-1)^\text{th}$  step,
so that $x(\dtrd)=rx_0$ for some $r\in[0,1]$.
We refrain from setting $r=0$ when $x$ is large because if $x$ changes too much in the time span $\dtrd$ our approximation of constant $A$ will break down.
Since all we can do is change the value of $\tilde{p}_n$, the equality $x(\dtrd)=rx_0$ won't in general be
achievable.
We therefore instead minimize the quantity
\begin{equation}
\beta(\tilde{p}_n) = \| P(x(\dtrd)-rx_0) \|_2
\end{equation}
where $P$ is a positive semi-definite matrix which relates the importance of minimizing certain state variables over others.
This quantity can be rewritten as
\begin{align}
\beta(\tilde{p}_n) &= \left\lVert w-\tilde{p}_n v \right\lVert_2 \nonumber \\
w &= P\left[ (e^{tA} - r\I)x_0 
		+ \tilde{p}_{n-1} \left( A+\I/\tau_r \right)^{-1} \left( e^{tA}-e^{-t/\tau_r}\I \right)b
	\right] \nonumber \\
v &= P \left[
		\left( A+\I/\tau_r \right)^{-1} \left( e^{tA}-e^{-t/\tau_r}\I \right)
		- A^{-1}(e^{tA}-\I)
	\right]b
\end{align}
In this form it is clear that $\beta(\tilde{p}_n)$ is minimized when $\tilde{p}_n$ is chosen to be the complex projection amplitude of the vector $w$ onto $v$:
\begin{equation}
\tilde{p_n} = \frac{\langle v, w \rangle}{\langle v, v \rangle}.
\end{equation}

For reference, note that in the limit $\tau_r\rightarrow 0$, the vectors $v$ and $w$ simplify to
\begin{align}
w &= P (e^{tA} - r\I)x_0 \nonumber \\
v &= -P A^{-1}(e^{tA}-\I)b.
\end{align}

\section{Pseudocode for Modified GRAPE}

In this Section, we list our modifications to GRAPE for use with the non-linear resonator as
\alg{grape}.

\begin{algorithm}
\caption{Modified GRAPE algorithm.}
\label{alg:grape}
\begin{algorithmic}
\Require
    Target unitary $U$,
    target fidelity $\Phi_\target$,
    distortion operator $g$,
    initial pulse $\vec{p}_\init$,
    ringdown compensation steps $n_{\text{steps}}$,
    ringdown compensation step width $\tau_r$,
    ringdown compensation ratio $r \in [0, 1]$,
    [optional] list of samples $\{\vec{x}_i\}_{i=1}^n$.
\Ensure  Pulse $\vec{p}$ such that that $\Phi_g[\vec{p}] \ge \Phi_{\target}$,
    or $\overline{\Phi}_g[\vec{p}] \ge \Phi_{\target}$ if a list of samples is given.
\Function{Util}{$\vec{q}$, $\{\vec{x}_i\}_{i=1}^n$}
    \State \Return $\sum_{i=1}^n \Phi[\vec{q} | \vec{x}_i] / n$
\EndFunction
\Function{RingdownCompensate}{$\vec{q}$, $x_0$, $n_{\text{steps}}$, $\tau_r$, $r$}
	\For{$i_{\text{step}} \in \{1, \dots, n_{\text{steps}}\}$}
		\State $q_0 \gets$ last step in $\vec{q}$
		\State $w \gets P\left[ (e^{tA} - r\I)x_0 
		+ q_0 \left( A+\I/\tau_r \right)^{-1} \left( e^{tA}-e^{-t/\tau_r}\I \right)b
	\right]$
		\State $v \gets P \left[
		\left( A+\I/\tau_r \right)^{-1} \left( e^{tA}-e^{-t/\tau_r}\I \right)
		- A^{-1}(e^{tA}-\I)
	\right]b$
		\State append $\langle v, w \rangle / \langle v, v \rangle$ to $\vec{q}$
	\EndFor
	\State \Return $\vec{q}$
\EndFunction
\Function{FindPulse}{$U$, $\Phi_\target$, $g$, $\vec{p}$, $n_{\text{steps}}$, $\tau_r$, $r$\,[, $\{\vec{x}_i\}_{i=1}^n$]}
    \If{distribution samples $\{\vec{x}_i\}_{i=1}^n$ are not given}
        \State $\{\vec{x}_i\} \gets \{\vec{0}\}$ \inlinecomment{Use a single sample if no samples are given.}
    \EndIf
    \State $\beta \gets 0$
    \State $\vec{d}' \gets 0$
    \State $\overline{g} \gets \sum_{i=1}^n g[\cdot | \vec{x}_i] / n$
    \State $J_g \gets J(\overline{g})$ \inlinecomment{Precalculate the Jacobian of the distortion operator $g$.}
    \State $u \gets 0$
    \While{$u \le \Phi_\target$}
    	\State $\vec{q}, x_0 \gets \overline{g}[\vec{p}]$ \inlinecomment{Distort the pulse, keeping the final state $x_0$ of the distortion.}
        \State $\vec{q} \gets \textsc{RingdownCompensate}\big(\vec{q}, x_0, n_{\text{steps}}, \tau_r, r\big)$ \inlinecomment{Compensate the pulse for energy removal.}
        \State $u \gets \textsc{Util}(\vec{q})$
        \State $\vec{d} \gets \sum_{i=1}^n \vec{\nabla}_{\vec{q}}{\Phi[\vec{q} | \vec{x}_i]} \cdot J_g$ \inlinecomment{Use \cite{khaneja_optimal_2005} to calculate $\vec{\nabla}_{\vec{q}}\Phi$.}
        \State $\Delta\vec{d} \gets \vec{d} - \vec{d}'$ \inlinecomment{Find the conjugate gradient direction.}
        \State $\beta \gets \max\{0, \vec{d} \cdot \Delta\vec{d} / \vec{d}'\cdot\vec{d}'\}$
        \State $\vec{s} \gets \vec{d} + \beta \vec{d}'$
        \State $\alpha = \argmax_\alpha \textsc{Util}(\overline{g}[\vec{p} + \alpha \vec{s}], \{\vec{x}_i\})$ \inlinecomment{Perform a line search in the ``good'' direction.}
        \State $\vec{p} \gets \vec{p} + \alpha \vec{s}$ \inlinecomment{Update the pulse by the step $\alpha \vec{s}$.}
        \State $\vec{d}' \gets \vec{d}$ \inlinecomment{Set the previous gradient to the current and prepare for the next iteration.}
    \EndWhile
    \State \Return $\vec{p}$
\EndFunction
\end{algorithmic}
\end{algorithm}


\section{Static Parameter Distributions}
\label{sec:static-param-dist}

We discuss a well known and somewhat trivial modification to the GRAPE algorithm
that deals with uncertainties in physical parameters which are static with
respect to the length of a single shot of the experiment, whether the single
shot be spacial or temporal.
The classic example is a small inhomogeneity in the static field of an NMR magnet;
up to diffusion, the molecules stay fixed in space and therefore each spin
will have a slightly different resonance frequency corresponding to the
value of the magnetic field at its position.

If the distortion or Hamiltonian is not known precisely, but instead
follows a distribution, we can consider the 
\emph{conditional} performance function,
$\Phi_g[\vec{p} | \vec{x}, \{H\}_{i=0}^L] = \Phi(g(\vec{p}, \vec{x}) | \{H\}_{i=0}^L)$,
where we have allowed $g$ to be a function of an additional vector
$\vec{x}$ and made the dependence of $\Phi$ on the system and control
Hamiltonians explicit \cite{khaneja_optimal_2005}. 
We are then interested in maximizing the
marginalized objective function,
$\overline{\Phi}_g[\vec{p}] := \expect_{\vec{x}, \{H\}_{i=0}^L}[\Phi_g[\vec{p} | \vec{x}, \{H\}_{i=0}^L]$.
Since the expectation operator is linear, this implies that we
can find the gradients of the marginalized objective function by
averaging over the gradients of the conditional objective function,
$\nabla_{\vec{p}}(\overline{\Phi}_g) = \expect[\nabla_{g(\vec{p})}(\Phi(g(\vec{p}, \vec{x}) | \{H\}_{i=0}^L) \cdot J_{\vec{p}}(g, \vec{x})]$.
Numerically, it is convenient to approximate this expectation value
by maintaining a list of hypothesis about $\vec{x}$ and $\{H\}_{i=0}^L$.

\end{document}